\documentclass[conference]{IEEEtran}
\IEEEoverridecommandlockouts

\usepackage{comment}
\usepackage{amsmath}
\usepackage{cite}
\usepackage{amsmath,amssymb,amsfonts}
\usepackage{algorithmic}
\usepackage{graphicx}
\usepackage{tabularx}
\usepackage{enumerate}
\usepackage{textcomp}
\usepackage{xcolor}
\def\BibTeX{{\rm B\kern-.05em{\sc i\kern-.025em b}\kern-.08em
    T\kern-.1667em\lower.7ex\hbox{E}\kern-.125emX}}

\usepackage{url}
  
\begin{document}

\title{From Talent Shortage to Workforce Excellence in the CHIPS Act Era: Harnessing Industry 4.0 Paradigms for a Sustainable Future in Domestic Chip Production\\
}

\author{
    \IEEEauthorblockN{Aida Damanpak Rizi\IEEEauthorrefmark{1}, Antika Roy\IEEEauthorrefmark{1}, Rouhan Noor\IEEEauthorrefmark{1}, Hyo Kang\IEEEauthorrefmark{1}, Nitin Varshney\IEEEauthorrefmark{1}, Katja Jacob\IEEEauthorrefmark{2}, \\Sindia Rivera-Jimenez\IEEEauthorrefmark{1}, Nathan Edwards\IEEEauthorrefmark{3}, Volker J. Sorger\IEEEauthorrefmark{1}, Hamed Dalir\IEEEauthorrefmark{1}, and Navid Asadizanjani\IEEEauthorrefmark{1}}
    \IEEEauthorblockA{\IEEEauthorrefmark{1}University of Florida, Gainesville, FL, USA}
    \IEEEauthorrefmark{2}{ZEISS Microscopy, Jena, Germany}
    \\\IEEEauthorrefmark{3}{US Partnership for Assured Electronics (USPAE)}
    \IEEEauthorblockA{Corresponding emails: adamanpak@ufl.edu, nasadi@ufl.edu}
}

\maketitle

\begin{abstract}
The CHIPS Act is driving the U.S. towards a self-sustainable future in domestic chip production. Decades of outsourced manufacturing, assembly, testing, and packaging has diminished the workforce ecosystem, imposing major limitations on semiconductor companies racing to build new fabrication sites as part of the CHIPS Act. In response, a systemic alliance between academic institutions, the industry, government, various consortiums, and organizations has emerged to establish a pipeline to educate and onboard the next generation of talent. Establishing a stable and continuous flow of talent requires significant time investments and comes with no guarantees, particularly factoring in the low workplace desirability in current fabrication houses for U.S workforce. This paper will explore the feasibility of two paradigms of Industry 4.0, automation and Augmented Reality(AR)/Virtual Reality(VR), to complement ongoing workforce development efforts and optimize workplace desirability by catalyzing core manufacturing processes and effectively enhancing the education, onboarding, and professional realms—all with promising capabilities amid the ongoing talent shortage and trajectory towards advanced packaging.
\end{abstract}

\begin{IEEEkeywords}
semiconductors, advanced packaging, industry 4.0, CHIPS Act, automation, augmented reality, virtual reality, workforce development
\end{IEEEkeywords}

\section{Introduction}\label{sec1:introduction}
The advancement of semiconductor technology, design, and performance is the cornerstone for transformations across all facets of the high-tech market, including high-performance computing, 5G telecommunication, Internet of Things(IoT), Artificial Intelligence(AI)/Machine Learning(ML) applications, aerospace/defense, and automotive electrification. The global sale of semiconductor chips hit an astonishing record of \$602 billion in 2022, as evident by the increasing digitization of industries around the world \cite{semiconductor_report}. Historically, the U.S. has been leading in the microelectronics revolution, starting from the advent of the transistor in 1947 to the new era of nanoelectronics—accelerated through Moore’s Law by the shift to FinFet and nano-ribbon technology. The progression of semiconductor manufacturing has allowed for substantial chip functionality, such as heterogenous integration in advanced packaging techniques of 2.5D and 3D chiplet integration for maximum chip-to-chip communication, which ultimately combines various functionalities into one package for a smaller device footprint. The myriad of rapid innovations necessitates more construction and operation of chip Fabrication Facilities (fabs). Since building advanced fabs accrues around \$20 billion in construction and chip manufacturing presents significant complexities, only Integrated Device Manufacturer companies (IDMs) and foundries within the US are able to fabricate their own chips, so majority of U.S. semiconductor firms maintain a fabless model. To keep up the pace, majority of U.S.-based semiconductor companies outsource their front-end and back-end manufacturing processing to overseas foundries for wafer fabrication and Outsourced Semiconductor Assembly and Testing (OSATs) for Assembly/Testing/Packaging (ATP). The lower construction and operating costs, along with presence of OSATs, drives companies to build fabs abroad; consequently, close to 80\% of fabs are based overseas \cite{fueling_american}. Similarly, at least 81\% of OSATs are also overseas, with Amkor and ASE being the major ATP powerhouses. According to Semiconductor Industry Association (SIA), the US semiconductor industry makes up 50\% of annual global market share, but constitutes only 12\% of the global manufacturing capacity. Furthermore, merely 6\% of development in manufacturing is located within the US. Consequently, the U.S. is generations behind the leading-edge chip manufacturers in Taiwan, South Korea, and China \cite{Shivakumar2022}. A confluence of factors, such as political conflicts and instabilities, national security concerns, as well as increasing supply chain vulnerabilities has driven policy makers to become wary of the geopolitical and economic threats associated with off-shored manufacturing in weakening the U.S. supply chain. In response, the U.S. Congress officially enacted the Creating Helpful Incentives to Produce Semiconductors (CHIPS) for America Act in October 2022 \cite{sia_2022}, which invests \$280 billion to bolster the research, development, and domestic fabrication/manufacturing of semiconductor chips in the U.S. Figure1 breaks down the CHIPS Act funding allocations, while Table1 provides the specific subsets allocated \cite{Badlam2022,chips_act2022}.

\begin{figure}[!hbp]
    \includegraphics[width=\linewidth]{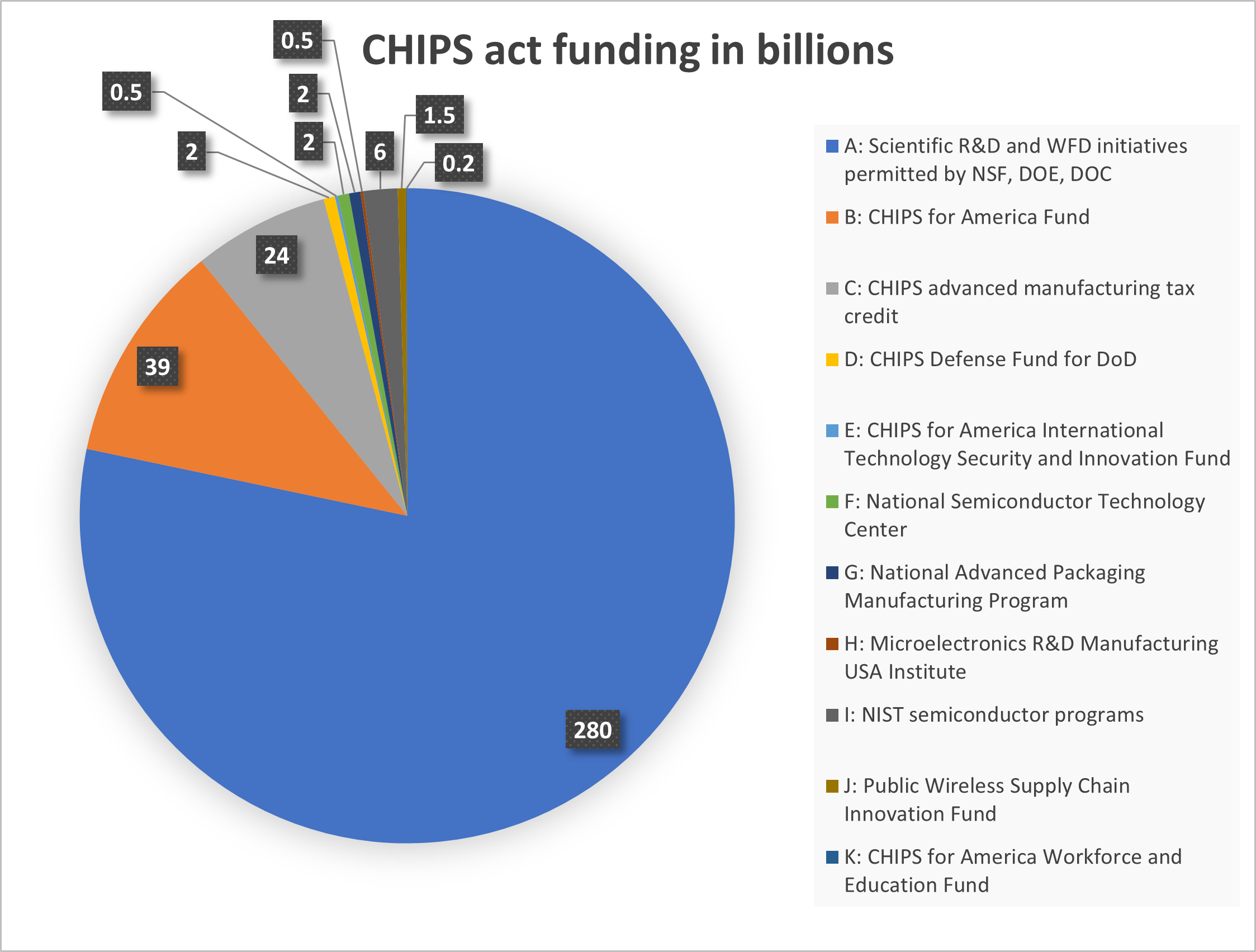}
    \caption{CHIPS ACT funding breakdown (in billions)}\label{fig:1}
\end{figure}

\begin{center}
\begin{table*}
\caption{CHIPS ACT Funding Allocations} \label{table-1}
\label{table:1}
{\renewcommand{\arraystretch}{2}%
\begin{tabularx}{0.99\textwidth}{|X|X|}
  \hline
  Subsets of CHIPS Funding & 
  Purpose\\
  \hline
  A: Scientific Research and Development(R\&D) and WFD initiatives permitted by National Science Foundation(NSF), Department of Energy(DOE), Department of Commerce(DOC) & 
  Dedicated towards funding research, workforce, and economic development programs\\
  \hline
  B: CHIPS for America Fund & 
  Construction and expansion of domestic manufacturing facilities\\
  \hline
  C: CHIPS advanced manufacturing tax credit & 
  A 25\% investment tax credit for advanced manufacturing and new facilities\\
  \hline
  D: CHIPS Defense Fund for Department of Defense(DoD) & 
  To implement the Microelectronics Commons, a network to onshore the prototyping and lab-to-fab transition of semiconductor technologies for WFD in advanced defense systems\\
  \hline
  E: CHIPS for America International Technology Security and Innovation Fund & 
  To support the development and adoption of secure telecommunications technologies and semiconductors\\
  \hline
  F: National Semiconductor Technology Center(NSTC) &
  To conduct advanced semiconductor manufacturing R\&D and prototyping for innovative technologies and advance workforce training\\
  \hline
  G: National Advanced Packaging Manufacturing Program & 
  A federal program dedicated to R\&D in advanced assembly, test, and packaging (ATP) capabilities through NSTC\\
  \hline
  H: Microelectronics R\&D Manufacturing USA Institute &
  Collaboration between industry, academia, and government to research virtualization of semiconductor machinery, optimizing ATP, and WFD\\
  \hline
  I: National Institute of standards and Technology(NIST) semiconductor programs &
  Various NIST programs to advance material characterization, instrumentation, testing, and manufacturing capabilities\\
  \hline
  J: Public Wireless Supply Chain Innovation Fund &
  To innovate the architecture of software-based wireless technologies, in the U.S. mobile broadband market\\
  \hline
  K: CHIPS for America Workforce and Education Fund &
  Initiate domestic WFD to resolve labor shortages through NSF\\ 
  \hline
\end{tabularx}}
\end{table*}
\end{center}

The U.S. is leading the market share globally in research and development (chip IP design) and equipment manufacturing \cite{fueling_american}. However, any effort by the CHIPS Act directed towards strengthening domestic manufacturing is undermined by the major talent shortage plaguing the entire spectrum, starting from a lack of technicians to design and operations engineers. This has alarming implications, as it keeps the U.S. from maintaining a secure manufacturing capacity in wafer fabrication and ATP, without a talented workforce driving it— which is paramount for innovating the next generation of semiconductor chips for advanced U.S. defense systems, automated machinery, as well as quantum computing. The key industry workforce development challenges that the CHIPS Act targets include \cite{fueling_american}:
\begin{itemize}    

    \item \textbf{Lack of brand awareness associated with semiconductor companies – } Semiconductor companies have a poor brand image and overall low recognition compared to other consumer-facing tech companies (hyper-scalers including Google, Microsoft, Amazon).

    \item \textbf{Industry going obsolete – }Decades of offshore manufacturing and fabrication has rendered this sector of the industry obsolete in the United States.

    \item \textbf{A lack of student interest in hardware electronics in comparison with software –} Starting as early as middle school, U.S. students have little to no exposure to basic electronics and hardware-oriented projects, a critical time for developing academic interests in exploring career opportunities. The focus is more on software skills, with computer science becoming one of the most popular college majors within Science, Technology, Engineering, and Mathematics(STEM) as students hope to find themselves working at big tech companies. As a result, majority of K-12 students remain unaware of the various technical opportunities that offer direct pathways to the semiconductor industry.
    
    \item \textbf{Outdated microelectronics curriculum – }As a result of misalignment between the high tech industry and education system, the current academic curriculum offer little to no exposure to modern semiconductor manufacturing and packaging techniques. This also includes K-12, where technical STEM exposure does not go beyond computer science fundamentals. Subsequently, there exists a wide gap between the entire talent spectrum, leaving a mismatch between K-12 STEM and college graduates’ skills and employer expectations within the industry .

    \item \textbf{Aging faculty and infrastructure – }A shortage of skilled faculty and outdated infrastructure calls for programs directed towards training instructors on state-of-the-art equipment for them to eventually integrate the gained knowledge and experience into their classrooms.
    
\end{itemize}

With \$39 billion dollars being funneled into funding new infrastructure, semiconductor companies are racing to build fabs, which ultimately creates thousands of job opportunities—however, the biggest hindrance will be finding workers to fill the gap with the onset of the talent shortage. There is a strong momentum to develop a secure semiconductor ecosystem in order to achieve long-term self-sufficiency in fabrication, manufacturing, and ATP—and the key in making this attainable is a stable and expanding workforce. While various new educational, training programs and stakeholder partnerships are targeted towards building the next talent pipeline, it is lengthy process with no guarantees. 

In this paper will explore the impacting parameters affecting semiconductor workforce development and proposes two Industry 4.0 capabilities: automation and AR/VR. It is structured as the following: section I. provides a high-level introduction to complex workforce issues that the CHIPS Act targets, as well as the motivation and opportunities from the enactment. Section II presents challenges the semiconductor industry is facing in establishing a domestic workforce and proposes Industry 4.0 solutions. A shift in technological paradigms in presented in section III, which illustrates the evolution of Industry 1.0 to 4.0 paradigms. Section IV introduces the Work Force Development (WFD) pipeline initiative model. Section V  discusses the challenges facing the WFD model as a result of fab working conditions.  The two technologies and future roadmaps are explored in section VI and VII through real-life use cases and unique research case studies conducted by previous literature. The ultimate goal is to showcase Industry 4.0 technologies as promising solutions in workforce education, training, and manufacturing to complement ongoing workforce development efforts in parallel and increase education/workplace desirability.

\section{Background and Motivation}\label{sec2:motivation}
\subsection{CHIPS Act and Workforce Development Challenges}\label{sec2_1:challenges_of_chips_act}
Semiconductor manufacturing encompasses a diverse set of expertise from various talent pools, including fab operators and quality technicians to electrical and materials science engineering PhDs performing specialized silicon design. A skills gap exists across the entire value chain within the industry. The main challenges the semiconductor industry is facing is explored, which is also summarized in Figure 2.

\begin{figure*}
    \centering
    \includegraphics[width=0.8\textwidth]{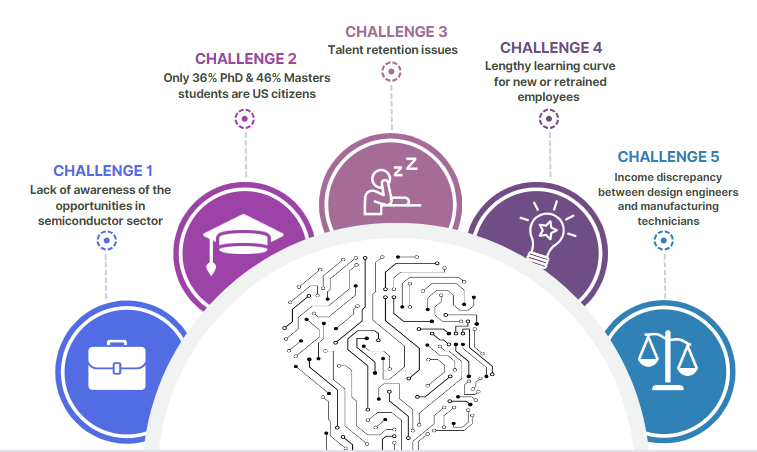}
    \caption{Challenges the semiconductor industry is facing in establishing a domestic workforce.}
    \label{fig:2}
\end{figure*}

\begin{enumerate}
\item \textbf{It is a career path that very few Americans are familiar with due to decades of outsourcing and doesn’t carry the same prestige as software jobs in big tech.} In a survey of 18 industry executives from various member companies of SEMI, 60\% felt that the entire semiconductor sector has a poor perception compared to other tech companies \cite{deloitte}, which tend to be more attractive in brand name. There is also a lack of awareness amongst employees in how dynamic the industry is, bearing the ability to switch to between roles as well upskilling for more lucrative opportunities. 

Even if the US merely focuses on maintaining domestic production for critical applications, this would entail up to 90,000 talented personnel required to staff new fabs \cite{Alam2023-hc}. Recruiting the right talent is a prevailing challenge despite extensive private and public investments towards reshoring semiconductor production. Furthermore, the U.S. has access to the largest talent pool across the globe, as it offers the best system of research institutions worldwide with endless opportunities. As a result, majority of international STEM students choose American colleges and universities to pursue their graduate degrees. 

\item \textbf{Subsequently, only 46\% and 36\% of engineering Master's and PhD students are U.S. citizens, respectively.} Limited number of domestic students end up pursuing advanced engineering degrees compared to international students, which spans a multitude of social and economic factors. For majority of domestic students, a Bachelor’s degree in any engineering discipline earned from U.S. universities is sufficient for a rewarding career, particularly when considering the time and capital investment spent towards graduate degree programs and a general lack of awareness about various scholarship opportunities to fund graduate programs.

To further accelerate the growth of the semiconductor 
ecosystem, there are three key strategies to remain
competitive in the talent war: reskilling, automation, and
expanding the pipeline through workforce development programs \cite{Alam2023-hc}. Leading Integrated Device Manufacturer(IDM) companies such as Intel, Micron, and TI have made notable strides towards reskilling for both current and future workforce through rotational and educational programs. Nonetheless, deploying all these efforts is a long-term investment that requires major time and capital to produce a substantial outcome, which makes it inefficient in accruing enough staff to operationalize the new fabs. Additionally, the mere investments in building fabrications plants will do little to offset the talent shortage, as it projected that by 2030, the U.S. will experience a shortage of 300,000 engineers and 90,000 technicians \cite{Alam2023-hc}. 

\item \textbf{Talent retention issues.} Many jobs within the industry, particularly fab roles, experience low employee retention due to impacting parameters affecting workplace desirability, such as lack of remote option, flexibility, salary, and compensation.

\item \textbf{To meet global chip demands for all market sectors, there is a lengthy learning curve for new or retrained employees to properly execute all the complex steps involved in the fabrication process within a fast-paced fab environment.} Even though large portion of semiconductor jobs are manufacturing intensive, majority remain unfilled with high attrition rates compared to design and analytic positions. 

\item \textbf{There is an income discrepancy between design engineers and quality engineers/manufacturing technicians working in fabs,} as the former is isolated from the hazards of working around risky machinery and chemicals as well as having the ability to maintain a flexible work schedule. The new generation of talent is placing a higher emphasis on hybrid/remote work options, especially after the 2020 pandemic. This is a concern of industry executives as fab roles happen to be less flexible than design counterparts and it is where the acute talent shortage is. However, the industry cannot expect their existing engineers and those in early stages of the value chain to give up better paying, lower stress, flexible desk jobs to be reskilled for demanding fab roles, which can further exasperate this issue as it leads to more attrition and dissatisfaction \cite{Alam2023-hc}.

\end{enumerate}

\subsection{Harnessing Industry 4.0 to Tackle Challenges }\label{sec2_2:harnessing_industry_4.0_to_tackle_challenges}
Ultimately, leading-edge innovations in Integrated Circuit(IC) chip packaging increases manufacturing complexity, which necessitates more process automation along with expanding the US capacity to hire more workers familiar with automation tools amid a competitive talent war. This specifically applies to technicians who are responsible for monitoring equipment and running recalibrations, especially with advancements towards microscale tool operations. The high capital investment that goes into recruiting, training, onboarding, and retaining employees requires dedicated training tools/programs and automation of repetitive tasks as a suitable option to re-structure the workforce, as it can alleviate the demands for scarce talent in the ongoing competitive war while maximizing workplace desirability. Traditional levers of automation include Automated Material Handling System (AMHS), and advanced dispatching and scheduling, which are older automation mechanisms commonly used in semiconductor fabrication facilities (fabs) processes. Moreover, the increased design and packaging complexities, along with outsourced fabrication and ATP, leaves the chip in immense vulnerability, creating the opportunity for potential adversaries to perform malicious activities by tampering with the chips, such as through Trojan insertion. This presents a need for rigorous testing, assurance, and inspection standards to detect and classify defects and Trojans—which can be a mostly automated process if embedded properly. Since a major target of the CHIPS Act is on chip design and wafer fabrication, automation in general can also garner support for increased funding and investment towards a domestic advanced packaging ecosystem since it reduces labor rates, increases supply chain security, and stimulates economic gains of equipment manufacturers by establishing ATP facilities \cite{Ver_Wey2022-qv}. However, the entire value-chain, especially backend manufacturing, have yet to reap the full benefits of Industry 4.0’s digitization for streamlined manufacturing/ATP. 

AR/VR is another Industry 4.0 technology that is sporadically used in the industry, which is becoming more accessible and widespread in various fields. The users are immersed within a digital environment that provides them with a real-world sense of simulated realms, allowing them to see and work with digital objects. The application of AR/VR has a transformative capacity in semiconductor education, workforce training/onboarding, and maintenance/manufacturing/assembly.

\subsection{Introducing the Roadmap to Automation \& AR/VR}\label{sec2_3:roadmap_automation_VR_AR}

The intricate chip design and fabrication flow of advanced packaging and technology necessitates a talented interdisciplinary workforce with extensive knowledge, skills, and abilities in various STEM domains. Nonetheless, U.S. community colleges, technical/vocational schools, and colleges/universities are not supplying enough talent to meet industry demands. Current estimates predict that workforce needs will double as the CHIPS Act, both directly and indirectly, generates thousands of job opportunities within the next couple years \cite{sia_2022}, but there is not enough talent to fill them.

The promising capabilities of automation and AR/VR, especially when used in parallel, increases workplace desirability, decreases manufacturing defects, and optimizes defect detection through recent advancements in AI/ML, ultimately maximizing yields in production.  Villani et al. \cite{Villani2021TheIS} provides a high-level analysis of industrial automation through case studies, which briefly demonstrates that despite rapid progression in modern production systems, human workers remain critical in industrial workplaces and that automation features as a complementing element by handling complex machinery operations to reduce heavy workloads on human workers. Leveraging automation mechanisms serves to improve desirability in work environments as it streamlines tedious and repetitive tasks.

\section{Industry 4.0 Applications in the Semiconductor Industry}\label{sec3:progression}
    The Industrial Revolution has shifted from mechanizing production to narrowing the gap between AI and brain power, through emulation of human thinking/decision-making by a computer. As highlighted throughput this section and Figure 3, each period has advanced by pushing scientific and technological limits to extend beyond human capabilities.
\subsection{Industry 1.0}\label{sec3_1:industry_1}
The roots of Industry 4.0 stems from the Industrial Revolution, which was propelled by the rapid development in science and technology in the 18\textsuperscript{th} century to streamline production and transportation through mechanization, with the emergence of the steam engine and the spinning wheel. Before Industry 1.0, iron production was arduous, which required workers to manually heat the iron ores with charcoal through a furnace. Eventually, this led to a timber shortage and called for a more efficient process and accessible fuel source. The discovery of coal revolutionized the industry as it fueled both the steam engine to provide mechanical power and revamped iron production \cite{industrial_rev}. By controlling the carbon content present in iron during smelting, the Bessemer process enabled the mass production of steel, which had superior properties to the brittle wrought iron. This transformed the global economy and infrastructure through the development of industrial machinery and complex railroad systems.

\begin{figure}
    \includegraphics[width=\linewidth]{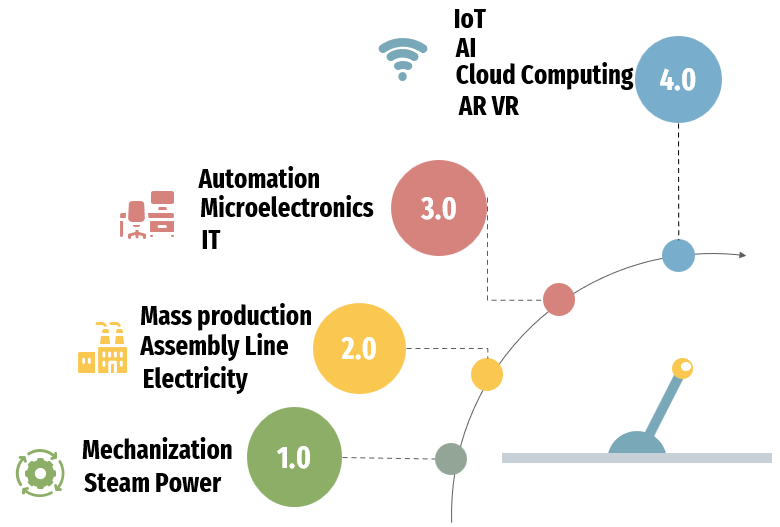}
    \caption{Industrial Evolution}\label{fig:3}
\end{figure}

\subsection{Industry 2.0}\label{sec3_2:industry_2}
The growth of the steel industry paved the way for Industry 2.0, which further modernized industrialization through electricity, assembly lines, and conveyor belts. Henry Ford pioneered the idea of mass production through the innovative combination of the assembly line to produce his affordable Ford Model T vehicles, which ushered in rapid technological globalization by revolutionizing transportation and general manufacturing \cite{model_2020}. Before his unique invention, only one station sustained the full assembly, which then transitioned to a distributed labor system divided into organized steps, inspired by the conveyor belts at that time. This incited the development of industrial-scale assembly lines, which deployed a novel systematic method composed of a rotating belt that transported parts for each worker to install, leading to an equal division of labor. As a result, the industry experienced increased worker productivity and decreased costs, ultimately simplifying and maximizing through-put for factories.

The key hallmark of Industry 1.0 was ``steam-power”, which harnessed the potential of mechanization to replace the ``muscle-power” endured from manual labor \cite{Wan2022TheRT}, while Industry 2.0 realized the colossal power of sophisticated machinery and electricity in further bolstering the economy.

\begin{figure*}
    \centering
    \includegraphics[width=0.85\textwidth]{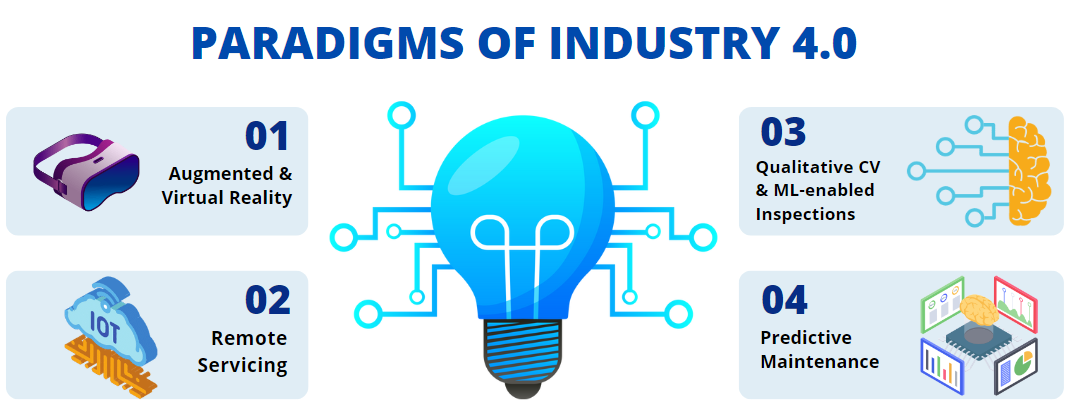}
    \caption{Applying the Industry 4.0 Paradigm in the Semiconductor Industry}
    \label{fig:4}
\end{figure*}

\subsection{Industry 3.0 evolving towards Industry 4.0}\label{sec3_3:industry_3_to_4}
In contrast, Industry 3.0’s main pivot was directed towards improvising production systems through microelectronics and Information Technology(IT), therefore introducing industrial digitization with advent of the computer. The digitization of factories is rooted in embedded programmable logic controllers incorporated into machinery to allow partial automation and streamlined data collection \cite{industry_4}. Industry 3.0 proliferated the power of automation via the rapid evolution of technological systems, as it currently exists in various aspects of manufacturing, including material tracking, equipment control, and product flow management to advanced process control. The benefits of automated metrology, decision-making, and analysis extends onto faster time-to-data, increased productivity and throughput, as it presents in Industry 4.0.

\subsection{Industry 4.0’s Unique Paradigms}\label{sec3_4:industry_4}
Industry 4.0 is characterized by full automation, intelligent tools with decision-making power, and big data analytics \cite{Butte2016-ii} to efficiently maximize productivity across the entire value chain. It is an amalgamation of recent innovations, such as: IoT, collaborative robots, digital twin modeling, AI, AR/VR, and cloud computing \cite{Zhou2015Industry4T, Cemernek2017-cz}. The seamless cooperation between each domain creates a reconfigurable and adaptive cyber-physical realm that aims to digitize manufacturing. Wan et al. \cite{Wan2022TheRT} defines Industry 4.0 as a fusion of three prime technological domains aimed at developing ingenious dynamics within modern manufacturing, known as the CIOT collaboration. The domains are Collaboration Technology (CT), Information Technology (IT), and the Operational Technology (OT)\cite{Wan2022TheRT}. 

\begin{itemize}
   \item \textbf{Collaboration Domain - }The exchange of big data is conceded through the CT domain, as it encompasses a wide range of communication protocols, including wireless networks such as Bluetooth and Zigbee as well as high-caliber cellular networks \cite{Wan2022TheRT}. 
     \item \textbf{Information Technology Domain - }To provide an intricate classification of OT data, the IT framework addresses the enhanced processing of complex datasets collected by humans and smart sensors through the aid of digital twin modeling, IoT, big data analytics, cloud computing, and artificial intelligence. With the current trajectory towards advanced manufacturing, traditional data analysis and synthesis techniques are projected to be overwhelmed by the enormous datasets that come from complex industrial processes \cite{Kumar2019-mt}, such as in 300 mm wafer fabs, especially as time progresses with growing production demands. 
   \item \textbf{Operational Technology Domain - }The OT serves to create a sustainable industrial operation/production ecosystem through effective management for maintenance of factories and creates a simpler interaction between employees and industrial equipment/materials.

\end{itemize}
\subsection{Industry 4.0 in the Semiconductor Industry}\label{sec3_5:industry_4_semiconductor_industry}
\textbf{All three ubiquitous domains form the basis for four potential innovative solutions applicable towards the semiconductor industry:} AR/VR, remote servicing, Computer Vision(CV)/ML enabled automated inspections, and predictive maintenance, as illustrated in Figure 4.
\begin{enumerate}
    \item \textbf{AR/VR }for remote planning and digital twin modeling. Employees can congregate through the CT domain to layout plans through AR/VR and visualize ideas, as well as exchange assistance. Furthermore, AR/VR can be used to simulate the 3D digital twin model of various fab scenarios for optimized production and reduce delays in congregating collaborative meetings through a virtual setting.
    \item \textbf{Remote servicing }is functional throughout all industrial settings. In midst of natural disasters or travel barriers, AR/VR can aid in remote servicing, where a group of experts can join to combat issues, such as equipment failure. It facilitates real-time interaction between the equipment manufacturers and clients by overcoming geographical barriers.
    \item \textbf{Predictive Maintenance }enabled through a network of IoT sensors. The data captured through the sensors is processed through mechanisms of big data analytics. ML algorithms can be developed to analyze the data, which workers can monitor to stay attentive of probable equipment failures that can halt production and create intervention plans ahead of time. With the tool to tool communications, there is not just predictive maintenance, but also  quality bound optimization and automated control adjustments between processes, such as course correction versus the compilation of tolerance failures during manufacturing. The aim is to reduce time spent by workers on diagnosing unexpected failures and repairs.
    \item \textbf{CV/ML enabled automated inspections  }to monitor production quality through analysis of visual data from the production line to spot possible defects and faults. The trained algorithms can streamline failure analysis to ensure maximum yield without needing manual quality inspection (which carries a higher chance of missed defects due to human error).
\end{enumerate}
Adoption of this integrated framework offers multiple unique venues that all play a crucial role in both front-end and back-end semiconductor manufacturing. Every century has carried innovations from all scientific domains to develop technology to supplement human productivity,which serves to optimize yield and maximize efficiency. The rest of this paper analyzes the feasibility of current WFD initiatives and its the pressing challenges, including talent retention and workplace desirability. 
\section{Current Workforce Development Initiatives: Shaping the Talent Pipeline}\label{sec4:workforce}
To build a secure semiconductor supply chain as a part of onshoring the U.S. semiconductor industry, a stable talented workforce is needed, especially as CHIPS Act fuels new facilities and fabs constructions across the country. A talented workforce is equipped with the skills to either design advanced chips through R\&D or through monitoring fab operations to ensure maximum production yield and quality, depending on which educational spectrum the individual is from. All of these factors call for urgent measure to establish WFD pipeline initiative that provides K-12 and college students with experiential learning opportunities through internships/apprenticeships, modernized academic curriculum with skilled instructors, and up-to-date research facilities and equipment. The amalgamation of degree programs offered by both community colleges and universities is best complemented with hands-on-experience to build and retain a resilient workforce \cite{fueling_american}.

This urgent talent gap has compelled extensive collaboration between various industry sectors, government, academic institutions, consortiums and organizations, to set the framework for the next workforce ecosystem, with the goal of creating diverse pathways to build and support a domestic workforce in microelectronics and advanced packaging to meet both immediate and future demands of the industry. The talent pipeline targets the entire educational continuum, including K-12 schools, community colleges, universities, and veterans—for a vast array of workforce development programs to reinforce industry awareness and train/onboard students to be well-prepared in their career endeavors. With the rise of the CHIPS Act, the new fabs being built will generates thousands of open positions for skilled technicians and engineers to fulfill. Companies are racing to partner with community colleges and universities to establish strategic partnerships in creating apprenticeship and internships programs across the entire academic spectrum. One notable case is Intel, with plans to build multiple fabs across Ohio, has promised to fund up to \$50 million to various institutions across Ohio that, other than sponsoring direct internship and research opportunities, will help to revamp the current curriculum, hire skilled faculty to teach advanced courses, and supply new lab equipment \cite{Patel2023_ref14}. Figure 5 provides a list of WFD initiatives dedicated to each education/skills domain, which are currently shaping the talent pipeline to establish defined pathway entries to the workforce.

\begin{figure*}
    \centering
    \includegraphics[width=0.99\textwidth]{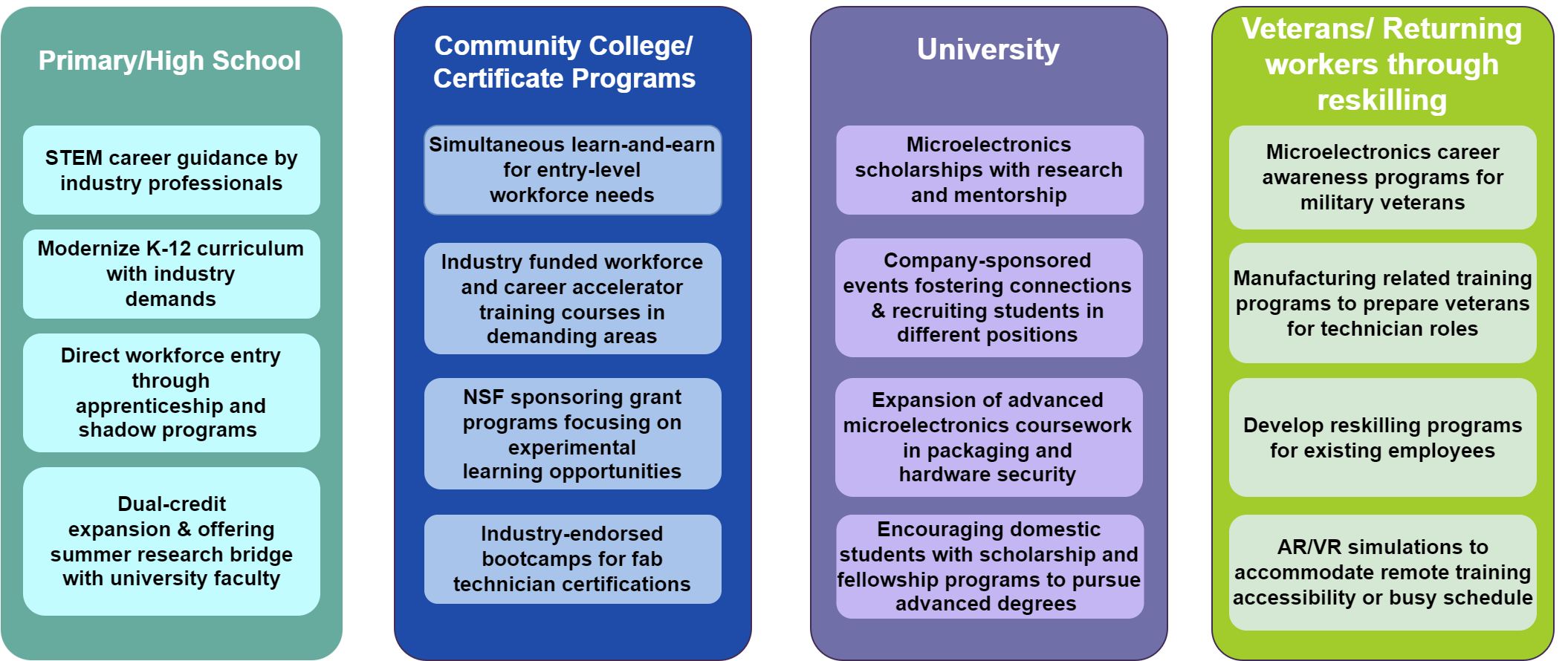}
    \caption{WFD initiatives to create a robust pipeline of talent – from core to advanced pathway entries to the workforce \cite{Zhou2015Industry4T,chapterMAPT,Kindling_undated-br}.}
    \label{fig:5}
\end{figure*}

\subsection{Initiatives within K-12 Education}\label{sec4_1:k_12_and_up}
Current downward trajectory of undergraduate retention rates within engineering programs has initiated substantial strategic efforts pivoted towards drawing pre-college students towards STEM, particularly electronics, as early as elementary and middle school. Since the average American school curriculum leaves out critical engineering disciplines with the exception of computer science courses, many talented students gravitate towards software, which leaves them alienated from the microelectronics field. As a response, companies and organizations are teaming with schools in taking remarkable initiatives to cultivate meaningful connections with K-12 students, such as STEM@GF, an online platform by Global Foundries (GF) that provides guided science projects for kids to develop core engineering skills \cite{STEM_GF}. The following initiatives introduce core hardware concepts, and familiarize students with careers opportunities \cite{Alam2023-hc}, \cite{Nathan_J_Edwards_Carter_Grizzle_Vaanathi_Sekar_Brett_Meadows_Michael_McGivern_Steven_Kiss_Asher_Edwards_John_Branning_Mohamed_Kassem2023-ec}:
\begin{itemize}
    \item {Gamified learning projects}
    \item {STEM outreach programs, summer camps, and extracurriculars sponsored by industrial partners}
    \item {Facility visits and invited professional as guest lecturers}
    \item {Mentorships in career exploration and research}
    \item {On-campus (high school) career fairs with industry visits to classrooms}
    \item {Industry-specific STEM camps or industry-led short courses with involved projects for direct workforce preparation}
\end{itemize}

Notable efforts are put forth by Samsung \cite{samsung}, which offers middle and high school outreach programs, known as the Semiconductor STEM Academy Experience, dedicated to exposing 6-12\textsuperscript{th} grade students to careers path in the semiconductor and manufacturing industry. Either students themselves or the school district can request a tour of the facility and time with subject matter experts such as the engineers/technicians or inviting Samsung for school events and presentations. Micron offers similar initiates such as the Chip Camp in select U.S. middle school districts, where students get to explore key processing steps in chip production (processing silicon wafers, photolithography, etching, and ion implanting), various STEM activities, as well as receiving mentorship by both Micron team members and engineering intern students. Notable student group projects include designing a solid-state drive using Lego parts under multiple constraints to stimulate innate creativity. For high school graduates, Samsung provides high school graduates a 10-week internship in a technician career track, enabling them experience working in the industry. The first two weeks involve a mentor guiding the interns to help them determine if this is a suitable career choice for them. The final week consists of a showcase event with intern presentations, and select students will be invited to transition into their Fab Apprentice Program, an earn-as-you-learn initiative. This program allows the aspiring technicians the opportunity to work with 2 days a week while completing their associates degree in engineering technology or manufacturing, and upon graduation, they will be eligible to receive a full-time technician role at Samsung. Engineering students also have the opportunity to intern as a semiconductor engineering intern, a 3-month internship they are assigned a project in an array of areas such as analytics, automation, diffusion, etch, photolithography, etc. with the opportunity to transition to a full-time engineering position as well.

\subsection{Continuing Technical Education Programs}\label{sec4_2:continue_teched}
A particular focus of WFD programs also consist of boot-camp courses offered by various community colleges to train future technicians, such as the community college system of Arizona \cite{Maricopa_Corporate_College_undated-zr}, one of the key states that is projected to become the next technological hub with the ongoing construction of fabs by both Taiwan Semiconductor Manufacturing Company(TSMC) and Intel. The routine operation of these fabs relies on the hands of clean-room technicians, who facilitate production manufacturing processes, oversee the quality and flow of equipment functions, and handle general maintenance. Candidates don't require advanced degrees to take on these jobs, merely needing a short-term training course to meet the position qualifications. Such programs are targeted towards skilling and hiring equipment maintenance technicians. Particularly, this includes graduates of technical Associate degree programs, veterans entering the workforce, and skilled workers from other sectors of manufacturing (with transferable skills), as all are credible sources of talent in for such positions.

A case is The Registered Apprenticeship at GF, an apprenticeship program that utilizes the learn-and-earn approach through hands-on training and classroom instructions for full-time technician roles. The training is free, targeted for all backgrounds with a high school diploma. Additionally, program graduates have the option to pursue an Associate degree afterwards. Through this program, all three stakeholders benefit within the partnership: the industry receives a constant flow of employees, each program graduate gets immediate employment with new skills, and the training program receives a pipeline of student participants to optimize new content delivery modalities for adult learners \cite{McCaughey2023RegisteredAF}.

To further expand educational and training initiatives, Purdue University established the Semiconductor Degrees Program's interdisciplinary track to offer undergraduate, masters', and postgraduate courses that encompass all key steps in chip production, such as chemical and materials processing, manufacturing, packaging, and supply chain management, as well as associates degrees through the nearby community college. 

In brief, the CHIPS Act has compelled key pathway agreements between 2-year and 4-year institutions to increase the flow of student transfers, new multidisciplinary educational and training programs (internships/apprenticeships) to expand and diversify the workforce, and strategic industry partnerships with leading semiconductor companies and organizations \cite{Weller2023-vi}.

\section{Challenges to Development of a Secure Workforce}\label{sec5:challenges}
However, while the WFD talent pipeline initiative has a promising vision on securing and expanding a robust workforce, significant time commitments and the current working conditions within the industry bring forth major challenges to onboard and retain a stable workforce, particularly for existing fab roles. Successful recruitment and hire of new technicians and engineers may not securely close the existing gap if the impacting parameters that affect workplace desirability of microelectronics jobs are not addressed, as a high attrition rate remains a major issue.

\begin{figure*}
    \centering
    \includegraphics[width=0.85\textwidth]{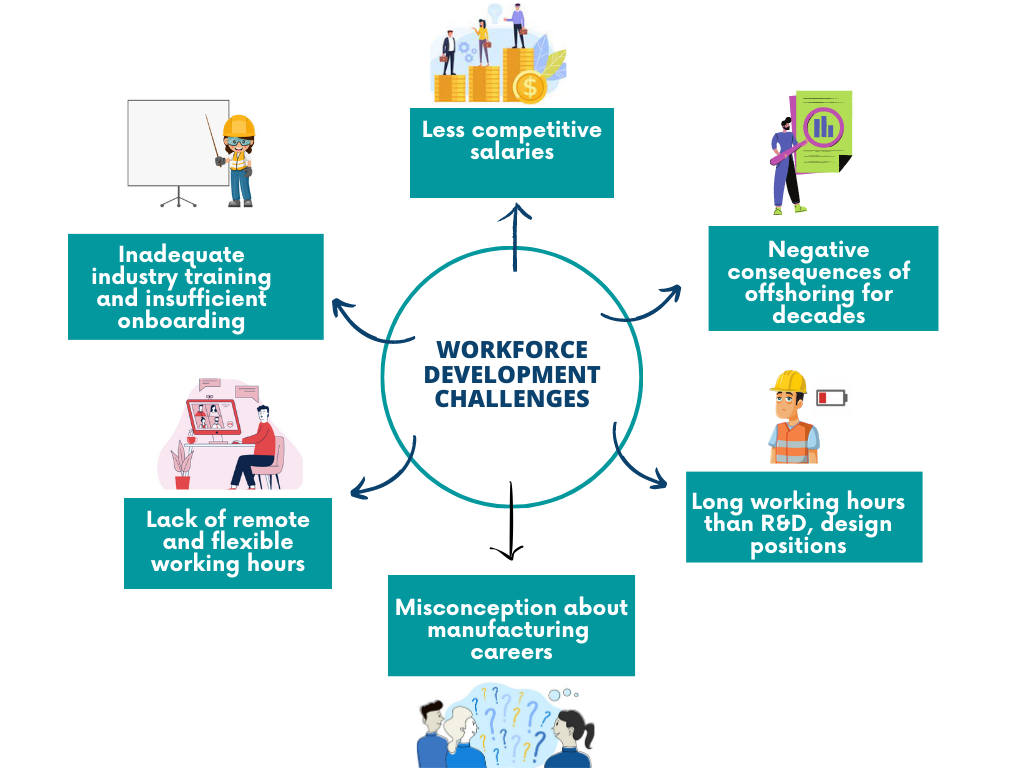}
    \caption{Key industry workforce development challenges}
    \label{fig:6}
\end{figure*}

Since majority of critical jobs are manufacturing-intensive, there is a lack of talent inflow into those roles. 
The various parameters, which are also highlighted in Figure 6, include:
\begin{enumerate} 
    \item Long overtime hours spent in fabs
    \item Lack of industry preparation and limited onboarding
    \item Less competitive salaries compared to other sectors
   \item Lack of remote options within manufacturing roles
\end{enumerate}
All of these synergistically contribute to a work-life imbalance. As the aging workforce retires, recent college graduates view manufacturing roles as reserved for those that did not pursue a college degree, as they also hold contrasting perceptions compared to previous generations, such as placing higher emphasis on flexible working hours to fit their lifestyles \cite{Deichler2021-gv}. Having the option to work remotely from home is another motivating factor that’s not feasible in the manufacturing realm. Within specific engineering roles at semiconductor companies, it is common for process, quality, or equipment engineers and technicians operating the fabs to endure prolonged working hours for weeks compared to those working in R\&D, design, and analyst positions in a corporate setting. This is highlighted with the multitudes of issues that are already starting to surface with the clash between U.S. employees and TSMC's working conditions. 

\subsection{Case Study: TSMC}\label{sec5_1:tsmc}
TSMC, a global leading manufacturer of semiconductor chips, is spending \$40 billion to build two major semiconductor foundries in Arizona, based on the 3nm and 4nm process, with plans to produce the most advanced chips in building a resilient U.S. supply chain that meets annual chip demands. To do so, it will need approximately 4,500 new hires to operate the two fabs, but has already earned a poor reputation amongst both current and prospective employees. On the company’s Glassdoor page, employees anonymously reveal their experiences about the brutal working environment: cultural clashes, high-stress fab work, misaligned management, 12-hour work days, and a cultural rift \cite{Lau2023-du}. Meanwhile, U.S. workers are accustomed to standard working conditions with expectations of a competitive salary and reasonable compensation. As a part of the onboarding process, U.S. staff are required to train at the company’s main hub in Taiwan for 12-18 months, which is a considerable downside for potential graduates entering the workforce. The next issue is salary, which is comparatively low compared to big tech and even rivaling companies in the industry, notably Intel, Micron, and GlobalFoundries—both of which still experience similar challenges in staffing their news plants despite a well-established reputation within the U.S. technological industry. 

Decades of offshoring has not only eroded the domestic supply chain and experienced workers, but the entire talent ecosystem and expertise as well, including instructors, consultants, and other support systems needed to sustain the semiconductor industry and even extends to general high tech manufacturing. Regardless of the current manufacturing gap and staffing challenges, doubled by the hurdles in initiating new plant operations, the potential geopolitical and economic gains far outweigh the risks for leading companies \cite{Lau2023-du}. 

\ To summarize, rebuilding a successful workforce with continuous output of talent while improving workplace desirability for the next generation is a part of CHIPS Act’s objectives coming to fruition. To keep U.S. competitive in the manufacturing race calls for measures in investing in two innovative solutions to complement WFD efforts: Automation and AR/VR. To revitalize domestic chip production, the semiconductor industry immensely benefits from leveraging digitization, realized through Deep Learning (DL) (a subset of ML), along with Computer Vision(CV), in broader areas of manufacturing, packaging, and failure analysis/detection \cite{Lee2021-lo, Nakazawa2019-ki, Kim2020-pw, Batool2021-yv, Lu2022-mt, Adly2014-nk, Hsu2021-qt, Ma2019-tv, Richter2019-pg, Ghosh2021-lg, cryptoeprint:2022/924, Asadizanjani2017-xe}. This is coupled with operation processes becoming more sophisticated with heterogeneous integration in the context of advanced packaging techniques. The following section introduces various opportunities in automation and AR/VR to streamline various fab roles to support worker productivity and maximize workplace desirability.

\section{Harnessing AR/VR Technologies in the Semiconductor Industry}\label{sec6:ar_vr}
    Both augmented and virtual reality are becoming growing areas of innovation across various sectors, with ubiquitous functionalities in the education, training, and professional space using VR head-mounted displays, as shown in detail in Figure 7. The main contrast of augmented reality is it combines the physical and virtual realms for a seamlessly interactive environment, in which simulated 3D objects are blended in the real world, while virtual reality is pure simulated replicas of the real-world. The rapid advances in both allows for an enhanced learning experience through realistic practice, which gives users a very high level of immersion and contributes to long-term learning sustainability. As a result, both AR/VR express a promising potential to revolutionize skill acquisition in all domains.

\paragraph {AR/VR in Learning \& Training}
\begin{itemize}
    \item \textbf{Engineering Education:} Both can also transform traditional education by enhancing student comprehension of core engineering subjects and improving performance through gamified simulations, done in parallel with experiential learning in physical laboratories.  Since current microelectronics training programs are faced with aging faculty and a lack of state-of-the-art equipment for lab operations and design software, AR/VR simulations be leveraged to provide a realistic view to clean-room and fabrication equipment through training modules, which is a limited ability for many programs with physical laboratories due to geographical barriers \cite{fueling_american}. Swansea University in Wales has realized this by launching a VR application that introduces the semiconductor industry to school children through gamification for an interactive and intuitive learning experience, as a part of its efforts to expand the industry regionally. The provided virtual interface takes the users through a virtual journey on clean room gowning, lithography, and highlights the importance role of semiconductors as the backbone of modern technology \cite{Thomas2023-ba}. To summarize, AR/VR increases interest in quality learning outside of a mundane environment and equips students with enhanced problem-solving skills to tackle real-world problems, since it gives them a chance to actively exercise their skills in a simulated world \cite{Makarova2015-dk}.
    
    \item \textbf{Employee Training:} AR/VR also supports training novice technicians in realistic work settings and maintenance, which accelerates knowledge acquisition and reduces repair operations (without relying on manuals only). Both serve as innovative training strategies to overcome skill barriers and provide flexible on-the-job training in fast-changing manufacturing environments \cite{Ulmer2020-xk}, which ultimately raises job satisfaction. This was realized through Ford Motor Company, where technicians were trained to diagnose and fix issues of the new e-vehicle's battery systems through modules using a VR headset. VR has the ability to attract new employees, which companies can leverage to brand themselves as high-tech and forward thinking, providing efficiency through accessibility \cite{Dearborn2020-mc}. Additionally, AR technology even shortens the learning curve in medical training by providing doctors with precise virtual surgical simulations. Without having to rely on surgical dummies, it allows doctors to practice anywhere and anytime through a VR headset and a haptic device \cite{Mileva2019-mz}. 
\end{itemize}

The demonstrated use cases illustrate the benefits and outcomes of using AR/VR technology, which can be applied to generate student interest in technical manufacturing as well as to train semiconductor technicians in complex tasks to increase employee engagement in interactive learning.
Interviews were conducted with Research Microscopy Solutions (RMS) experts to provide a Subject Matter Expert (SME) perspective on the use of VR within the industry, specifically for the microscopy community and service engineers. As shown in detail in Figure 8, the general consensus from both experts is that VR provides an immersive learning experience that is sustainable for long-term information retention, as it allows direct hands-on learning through a simulated but realistic environment. While it cannot completely replace traditional learning formats such as physical hands-on experience, it provides a complimentary learning mechanism to foster rapid learning/training with a realistic sense in scenarios that would otherwise be deemed unfeasible.

\paragraph {AR/VR Use Within the Industry}
Other than learning and training, it provides 3D prototyping and remote servicing. 3D-models are converted into Computer-Aided Design(CAD) data files as replicas to construct the virtual environment as a part of the advanced manufacturing mechanism. This allows companies to optimize product quality through enhanced design and modeling, as well as to decrease production costs and the design-to-assembly time\cite{Hamid2014VirtualRA}. Enabling the manipulation of the manufacturing models in the computer, helps users to be immersed in digital environment and improvise each step of the manufacturing process. Furthermore, the interactive environment supports the remote sharing of production data between workers, where one provides feedback and guidance on the operation flow via an established communication network that is attributed to the CT domain.

\begin{itemize}
    \item \textbf{Manufacturing/Assembly:} AR/VR are gaining significant momentum in the assembly and manufacturing spaces with the onset of Industry 4.0. Specifically, AR provides simulated path planning of robotics, maintenance, and assembly, as it allows workers to control quality during production with hands-free access. Spitzer et al. \cite{Spitzer2020-df} provides two cases for industrial uses of AR, one of which is an automated assembly manual system.  In this case study, AR is used to visualize changes during production, from the CAD engineer making design changes, to the assembly planners translating technical documents to instruction manuals and the shop-floor workers performing the updated assembly with it. The AR system adapts and visualizes any edits/updates made by the CAD engineer within the design, which eliminates the time-consuming process of making visual changes within the physical manual and cuts down production time. This increases worker productivity and job satisfaction as it allows the employees to make better decisions with a smoother operation, when the visualized information is embedded within the AR system. Botto et al. \cite{Botto2020AugmentedRF}  further supports AR for assembly use, in a  user study of 21 volunteers using AR on a tablet for assembling given models. The result was a reduced error rate during assembly, with a higher completion time, which is attributed to the limited background knowledge of the users in both AR and manufacturing.
    \item \textbf{Virtual Engineering:} Inspection and maintenance of equipment is one of the core activities in manufacturing. Remote inspection  approaches provide the opportunity to optimize the efficiency of maintenance processes without any time or geographical barriers or physical presence, as it allows technicians and engineers the flexibility to collaborate virtually from anywhere to perform the needed procedures \cite{Linn2017-uc}. 
Also known as virtual engineering, this concept was notably realized by ASML, a leading global supplier of lithography machines for semiconductor industry, with virtual engineers-in-charge. It unveiled an intuitive use of AR during the COVID-19 pandemic \cite{asml2020}, where by a task force of 100 SME across the world gathered remotely to service the lithography machines, which could’ve resulted in million-dollar implications for ASML and its customers in midst of the travel restrictions. The team rapidly developed an AR solution and after quick factory testing at the ASML site in Netherlands, the team fully embedded the AR platform on a smartphone as well as a Microsoft HoloLens headset to complete the necessary service actions at the customer site. The SMEs virtually entered the cleanrooms in customer fabs and completed the servicing via providing real-time instructions to the walk the on-site engineer through performing the troubleshooting and the actual service action— a innovative application of AR successfully applied to a dynamically complex scheme that was resolved through interactive communication converged on a virtual bridge between experts to accomplish the necessitated tasks. AR has immense capabilities in the instant deployment of expertise from anywhere across the globe to surmount any potential barriers.  
    
    \item \textbf{Digital Twin Modeling:} Another dynamic implementation is digital twin modeling for smart manufacturing. By recreating a physical system in a cyber realm, manufacturers can realize potential optimizations, achieve a greater increase in capacity, and maximize production efficiency without the risks associated in physical implementation, which was applied successfully to layout fab planning through the digital twin modeling and simulation of an automated 300mm Infineon Technologies fab \cite{Heinrich2012RulesOA}.
\end{itemize}

Nevertheless, widespread adaptation of AR/VR within both educational and industrial domains is a complex process. Addressing these limitations necessitates a comprehensive framework according to each domain's use standards, which is beyond the scope of this paper. 

\begin{figure}
    \includegraphics[width=\linewidth]{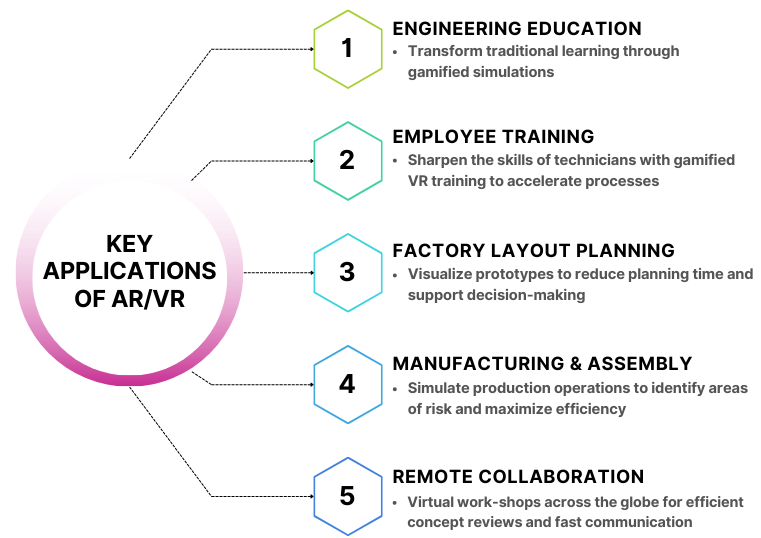}
    \caption{Applications of AR/VR}\label{fig:7}
\end{figure}

\begin{figure*}
    \centering
    \includegraphics[width=0.99\textwidth]{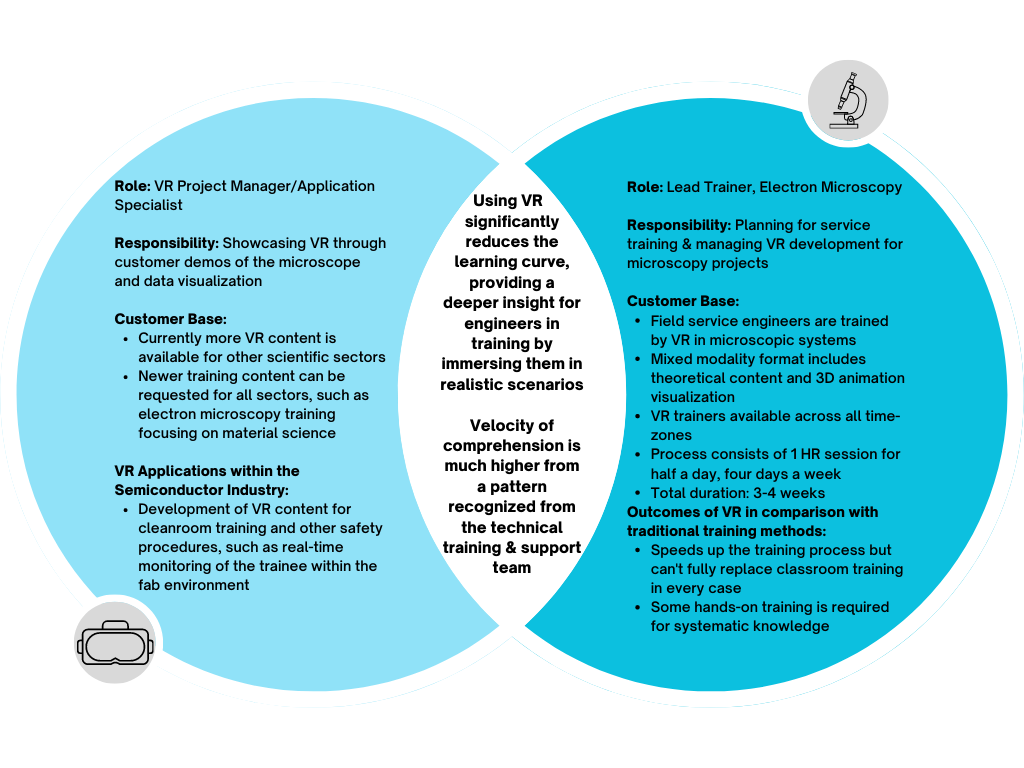}
    \caption{Industry perspective from experts on using VR for training}
    \label{fig:8}
\end{figure*}

\section{Automation: Current Case Studies and Future Road map}\label{sec7:case_studies}
\subsection{Current Automation Use Cases }\label{sec6_1:current_automation}

The process of semiconductor manufacturing is broadly split across three mains areas: 
wafer manufacturing, chip manufacturing (verification of internal circuitry), and product manufacturing (assembling and testing of the full chip) \cite{Liao2010-rz}. Automation within the semiconductor value-chain spans across a continuum, as the high-volume manufacturing environment is highly dynamic and errors of any degree within any step of production can drastically affect the quality and yield. 

Historically, front-end operations within 200mm fabs were manually handled by clean room technicians, such as material transportation, which affected production throughput due to more human error and higher risk of contamination during wafer processing—this was replaced by the AMHS, which reasonably reduced human action during manufacturing, as well as advanced process control, and production planning. As a result, current fabs are highly automated up to Industry 3.0 level.

Despite the evolved techniques of Industry 4.0, the semiconductor sectors, specifically back-end manufacturing ATP, do not fully harness the power of smart manufacturing, which restricts the valuable time of existing workers spent towards manual and repetitive tasks. Amid the talent shortage, automation releases the pressure the industry is facing by optimizing workplace desirability through automating repetitive and/or hazardous tasks and increasing worker productivity, as well as maximizing profitability in parallel, as automation tools streamline time-consuming processing, assembly, and quality control flow \cite{Alam2023-hc}.

 The SEMI Foundation \cite{Hanny_undated-fl} provides a spectrum that presents the progressing stages of automation for fabs, from level 0 (zero automation) to level 5 (full automation). Level 1 and 2 consists of Industry 3.0 techniques. For level 1, it consists of AMHS, defect control for yield tracking, and limited preventative maintenance, all of which aid the operator at a subsystem level. Advanced Process Control (APC) is introduced at level 2, which involves Run-to-Run (R2R) processing and Fault Defect Classification (FDC). Broader Industry 4.0 applications, such as AI/ML, are introduced in level 3 to provide conditional automation, in providing preventative maintenance techniques and digital twin modeling on cloud-based platforms. Level 4 and 5 further advance level 3 as with the move towards full automation, where AI/ML provides high-level cognitive abilities for operational decision-making.

In addition, back-end semiconductor manufacturing ATP, which is currently present across US-based foundries and IDMs, remains relatively labor-intensive as it has not fully embraced advanced automation technologies for its existing tasks. With the growing use of advanced packaging techniques, there is a call for more Industry 4.0 automation tools, since more manual tasks increase error rates and reduce worker productivity during the execution of complex procedures. The amount of time a worker spends handling various materials and machinery accounts for 30 to 50 percent of all labor, in addition to time spent idle for production cycle completion \cite{De_Backer2017-qd}.  

While the construction of new fabs across the U.S. will open up numerous potential jobs, Accenture presents that certain lower-level fab roles, extensively involved in laborious tasks, will be fading away by 2030 due to automation superseding manual labor across all segments of the industry, as forecasted by the U.S. Bureau of Labor \cite{Alam2023-hc}. A series of case studies will be presented that realize the potential in automating fab roles such as: equipment engineering and technician, quality engineering, characterization, reliability tests engineer, and process engineering. 
 
\subsection{Future Road map and Need for Automation }\label{sec7_2:cases}
Smart manufacturing, within the context of Industry 4.0, is a broad amalgamation of the following innovations and their role: IoT provides data collection platforms and communication mediums, cyber-physical systems programmed with predictive algorithms allow the physical and virtual systems to interact through a IoT platform which provides predictive equipment maintenance capabilities, and cloud computing serves as a network storage base for cloud manufacturing services such as product design and testing simulations \cite{Lin2017DevelopmentOA}. Large datasets of critical processing information is collected and analyzed by an ML algorithm to derive hidden patterns \cite{Khakifirooz2017BayesianIF})

 A particular use case within front-end manufacturing is Automated Visual Inspection (AVI) of wafers aims to reduce manual labor spent on quality control while increasing production yield. The general procedure of AVI consists of inspecting the device by sensors and processing the collected data through multitude of CV or ML techniques, which provides feedback throughout manufacturing flow for further optimization \cite{Huang2015AutomatedVI}. Any detected abnormalities notify the quality or process engineer in-charge for further fault assessment, only if a back-up Out-of-Control Action Plan is not implemented to automatically halt the operation to prevent potential yield loss \cite{Barar2020TakingEA}. As a result, this eliminated the need for manual data tracking and analysis.
\subsubsection{Wafer Quality Control} 
In regards to automating quality control during semiconductor manufacturing, Azamfar et al. \cite{Azamfar2020DeepLD} proposes an adaptable deep learning algorithm that’s applicable to a wide range of manufacturing settings that employ varying operating conditions, so it offers a generalized fault detection mechanism to monitor optimal wafer quality, as experimental dataset is obtained from actual semiconductor manufacturing. Using this method allows for predicting the quality of wafers without tedious manual inspection, ultimately reducing human intervention in quality control of fab processes. Since deep learning algorithms require supervised training using enormous datasets to ensure a well-trained model, the same prototype can’t be carried onto the manufacturing industry as it is not feasible to maintain such large database of labeled data that also accounts for varying operating conditions when considering the non-linear nature of manufacturing. This study overcomes this constraint by using cross-domain training for a domain-invariant Convolutional Neural Network (CNN) model, meaning that it’s applicable to various manufacturing processing conditions for reliable quality control. By using data obtained from multiple sensor sources, the trained model was tested with an unlabeled target data that’s also captured from various sensor sources. The model performance is evaluated using datasets obtained during real etching process in wafer processing, which produced favorable outcomes in demonstrating a generalized deep learning model that’s applicable for all scenarios in manufacturing.  The ultimate goal is to lower reliance on human expertise and lessen manual inspection, which also lowers risk of human error and increases worker efficiency by automating a repetitive task for process engineers \cite{Azamfar2020DeepLD}.

Defects occurring in semiconductor devices take on a variety of visual shapes and textures as the fabrication process is highly complex. As a result, manual inspection technique and classification remains dependent on the background of the experts performing the inspects. Imoto et al. \cite{Imoto2018ACT}  introduces a CNN-based transfer learning method for automatic defect classification that overcomes any ambiguity associated with manual classification, by assisting engineers in their tasks. The process consists of classifying images captured by the Scanning Electron Microscope (SEM) system, observing the frequency of each defect type and categorizing it for deeper failure analysis. Deep-learning CNN algorithms necessitate a large labeled dataset of training data, which is not feasible in large-scale manufacturing and various design IPs involved. Transfer learning serves as a solution to this issue through re-use of previous learned tasks on a limited database. At an actual manufacturing site, wafer surface SEM images of various defects were sampled for performance assessment. As a result, the labor for manual inspection was reduced by approximately 2/3 compared to commercially-used automatic defect classification methods.

\subsubsection{AI-Controlled Real-Time Cluster Tool Scheduler for Wafer Processing}
Cluster tools are an automated manufacturing mechanism that consists of multiple computer-controlled process units, a wafer-handling robot, and loadlocks (LLs) for loading and unloading wafers, as a part of the wafer fabrication process \cite{Pan2018-ua}. It allows for reducing the number of equipment needed in wafer processing steps by implementing various process chambers with the same recipe on one platform. Various wafer recipes rotating through the chamber on  different sequences, combined with the overall non-linearity of the manufacturing process, raises the processing complexity and inefficiency. Suerich et al. 
 \cite{Suerich2022ArtificialIF} deploys an AI-controlled cluster tool scheduler that actively collects real-time data and creates optimal scheduling algorithms, which are adaptable to any changes made in the tool processing times, without the need for human intervention. This was realized through virtual tests of various scheduling scenarios on the experimental digital twin models of the equipment (another Industry 4.0 technique) without risking machine downtime. The sensors and automation software is already embedded within the cluster tool system, where the AI algorithm is able to adapt the schedule based on any changes detected from collected data. The main limitation of this model is the time the AI spends searching for the most optimal scheduling solution, which calls for further improvising through offline optimization.

\subsubsection{IC \& System-Level Failure Analysis and Physical Assurance}
While front-end semiconductor manufacturing processes are in a better state in regards to automation, failure analysis and physical assurance of ICs and PCBs remain cases where the potential of intelligent mechanism is yet to be fully leveraged. Traditional inspections techniques have relied on human input to validate and check electronic components for both defects and malicious tampering, which is a time-consuming process prone to human error as the dataset becomes larger \cite{Zhao2022-ee}. As IC packaging opens the avenue for evolved systems through 2.5D and 3D chiplet integration, it also becomes more sophisticated and reliability becomes a much bigger priority, especially as the technologies gain widespread use in critical applications including aerospace, defense, and healthcare. The physical inspection techniques in validating IC packaging and PCBs are categorized into either destructive reverse engineering (mechanical/chemical cross-sectioning) or non-destructive methods through imaging (optical microscopy, x-ray, SEM). The ultimate goal is to streamline the detection and analysis of defects/faults using non-destructive methods as it is not sustainable for to utilize destructive reverse engineering methods in manufacturing setting. Particularly, as advanced IC packages possess highly intricate interconnections, it necessitates a comprehensive workflow with automated image acquisition, data extraction, and analysis to ensure structural reliability and streamline the repetitive task of technicians and engineers involved in manufacturing lines \cite{Yang2021-cp}.

Optical inspection remains the gold standard for detecting any external or internal abnormalities for quality control in part of minimizing yield loss, as it is versatile for all types of electric components without needing destructive sample preparation. Defect classification of cross-sectional images obtained from optical, x-ray, or SEM tools is an arduous task when done manually and necessitate use of virtual metrology and advanced defect classification algorithms. Additionally, miniaturized features in recent nodes may result in noisier images being produced, thus leaving failure analysis heavily dependent on SME input without extensive image processing techniques. However, rapid progressions in machine learning algorithms have allowed for image recognition to become an automated task \cite{Silva2019-bw}, particularly with the advances in ML, particularly CNN, a subset of Deep-Learning Neural Network (DNN), designed for processing high volumes of data. Since defect detection is highly complex process, CV solutions alone cannot provide full assurance. Subsequently, optical inspection solutions are scaling towards ML/DL methods for a reconfigurable approach to failure analysis, which also reduce the need for preprocessing the image \cite{Jessurun2022-vr, Yang2021-bz}. Both techniques can be combined with traditional CV algorithms to complement newer ML-based techniques by extracting parametric features to reduce the amount of data needed to achieve high accuracy when training ML models \cite{Zhao2022-ee}. This reduces the burden on fab engineers that utilize defect information to optimize production lines \cite{Peters2022-wd}.

\begin{itemize}
 \item \textbf{Automated Via Detection for PCBs:}
While this case study's main focus is on Printed Circuit Board (PCB) assurance, IC chips constitute a major part of any electronic system within a PCB, and both chips and PCBs happen to be the most counterfeited component as a result of outsourced manufacturing \cite{Sathiaseelan2021-zh}. Non-destructive reverse engineering of PCBs using X-ray computed tomography remains a 
manual process with a heavy reliance on SME presence, which requires comprehensive inspection to ensure system-level integrity of electronic devices. Botero et al. \cite{Botero2020-pl} explores the automation of non-destructive PCB reverse engineering by
presenting automatic detection of vias, which are small openings that connect the circuity of an IC device and vary a lot depending on
the size and imaging conditions. The via detection framework uses image processing techniques to first detect the via, using 3D constructed images captured from x-ray computed tomography, and then iterates through an unsupervised multi-step process to remove false-positive via detection. The results were confirmed using a deep learning CNN-based algorithm. The aim of the study was to present an automated non-destructive reverse engineering techniques that minimizes human     intervention for validation and verification of  hardware boards.
\end{itemize}

\begin{itemize}
    \item \textbf{Isolating IC TSV Defects:}
    \begin{itemize}
    \item Kim et al. \cite{Kim2018DeepLB} proposes a deep neural network (CNN) architecture to be leveraged for Through-Silicon Via (TSV) defect classification in 3D-ICs through comprehensive analysis in feature extraction and pattern identification. The methodology consisted of training the CNN model with a dataset of TSV defect images from x-ray microscopy and SEM, which outperformed traditional image processing techniques by reducing human dependance throughout the process by as much as 78.6\%, but nevertheless requires a sharper classification accuracy for size-dependent classification if it is to be deployed in fab testing and packaging processes.
    \item Wolz et al. \cite{Wolz2022XrayMA} presents a more detailed view into using x-ray microscopy and deep learning algorithms for precise isolation of wall delamination defects in micrometer-sized Cu-lined TSVs with a low depth-to-diameter ratio, to be used in quality control and process efficiency within R\&D settings. This study visualized the delamination defects by scanning both the whole IC device and the TSV through varying resolutions, with the latter providing better image quality down to sub-micron scale. To establish precise feature extraction and characterization, the database consisted of 30,000 objects to provide a fully automated defect analysis.
    \end{itemize}    

    \item \textbf{Detecting flip-chip C4 bump defects through SAM:}
    Scanning Acoustic Microscope (SAM) is a fast and non-destructive failure analysis technique to identify and inspect faults and isolate defects within microelectronics opaque internal structure, as well as ghost markings, using multiple types of material characterization scans to validate the integrity and authenticity of the chip \cite{Johnson2021-qd}. The transducer sends ultrasonic signals of a certain frequency into the sample, and any reflected signal is a result of changes in material density, voids/cracks, and delamination. The reflected signal is detected by the same transducer and converted back to an electric signal, which is combined to construct an image of the internal structure \cite{YazdanMehr2015AnOO}. Since a majority of automated classification methodologies use either X-ray, optical microscopy, and SEM images to form the image-based algorithms, a large set of high-quality data is required for training. Nair et al. \cite{Nair2022AutomatedDC} proposes a systematic deep-learning approach to interpret and analyze the detected ultrasonic signals autonomously using semi-supervised training with a limited labeled dataset. The model was tested to categorize flip-chip C4 bumps into defect and non-defect sets using the A-scan mode through a CNN algorithm, which outperformed other models in precise classification. 
    
    While further optimization of the current model’s accuracy and algorithm is needed to achieve full-scale industry use, this proposed model resulted in a favorable outcome in automating defect detection for maximum production yield through locating and eliminating a variety of faults while minimizing human expertise in signal interpretation.  
\end{itemize}

\subsubsection{Advanced Packaging Manufacturing}
    \begin{itemize}
    \item \textbf{Screening On-Package Faulty Passive Device:}
    Integrated Passive Devices (IPDs) have become increasingly prevalent in semiconductor advanced packaging, enhancing power integrity, and impedance matching. The demand for guaranteeing signal and power integrity in chips used in safety-critical applications like automotive, aviation, industrial, and defense systems has increased. IPD is used in various analog and Radio-Frequency Integrated Circuit(RFIC) for enabling on-package passive devices. IPDs significantly improve the quality and reliability of these chips and their packaging. As a result, conducting comprehensive testing and screening of IPDs becomes essential. It is crucial to note that replacing faulty IPDs far exceeds the manufacturing cost, emphasizing the significance of screening defective IPDs before their installation. Chuang et al. \cite{IPD_detection} proposed a machine learning-based screening method to detect faulty IPD, specifically capacitors with potential reliability issues. Using parametric data gathered from 360,000 integrated passive devices (IPDs) during the wafer probing test, their developed ML model’s primary objective is identifying IPDs with low breakdown voltage, which signifies reduced reliability. Using the ML algorithm, this method can successfully detect and eliminate 6 to 15 times more faulty dies, significantly improving the screening process. 
    Automating the detection of faulty on-package passive devices will result in an increased overall chip manufacturing yield rate. However, this works only focuses on detecting the capacitor. Further automation is needed to increase reliability to detect faulty on-package passive devices like inductors and resistors.

    \item \textbf{Optimizing Manufacturing Process for Reliability:}
    As advanced packaging is increasingly getting complex due to the introduction of 2.5D or 3D packaging, the reliability of semiconductor packaging is becoming an issue. Adapdix\cite{adapdix_manufacturing} developed an ML algorithm to optimize the manufacturing process to ensure reliability in semiconductor packaging manufacturing. An example showcases utilizing an AI/ML platform that operates directly at the machine's edge, leveraging real-time machine data and operational data. This platform integrates with various machine data sources such as sensors, actuators, PLC, and log data, utilizing appropriate communication protocols. This integration enables access to a wide range of real-time data, including parameters like acceleration, rotation, conveyor speed, gripper position, bonding head motion, epoxy dispense pressure, and other relevant machine part parameters. These real-time data points are critical in evaluating machine and process performance. This foundation facilitates the development of ML models, effectively harnessing the power of the collected data. Numerous prominent global enterprises have successfully employed these capabilities in semiconductor fabs, advanced packaging, fiber optics for transceivers, and precision placement for Surface Mount Technology (SMT). Remarkable enhancements have been observed, including improved alignment accuracy, increased process yield, reduced cycle time, and minimized machine downtime. Using automated data acquisition and analysis with AI/ML will effectively minimize the human expertise to detect and predicts faults manually. As semiconductor packaging continues evolving and becoming more complex, there are future scopes for the researcher to improve automation and optimization of semiconductor manufacturing to solve the reliability problem. 

    \item \textbf{Automated Thermal Characterization of IC Packaging:} Proper thermal characterization of packages  ensures maximum performance and reliability of ICs for power electronic applications \cite{noauthor_2007-vf}. To maintain the IC junction temperature at a optimal level, there needs to be effective thermal management of heat flow through the system's electrical paths, from the IC to the substrate. One of the roles of a packaging engineer is to predict and classify the patterns of IC package substrates' thermal conductivity for a reliable end device. However, this is becoming a labor-intensive task with potential human errors due to increasing complexities in advanced packaging designs. Kim et al. \cite{Kim2022-qm} proposes a CNN-based algorithm for an automated prediction algorithm that determines the Effective Thermal Conductivity (ETC) of packaging substrates using CNN for heat transfer optimization. The CNN algorithm is trained to recognize and extract patterns from the imaged substrate layers for precise prediction by dividing the layer-pattern images into single unit cells. From there, the local ETC is determined from each cell, with all of them re-grouped into a larger unit  afterwards for a comprehensive thermal analysis. This result were confirmed through finite element simulations of the package CAD file. The ETC serves as a critical parameter of temperature distribution since the IC constitutes the main heat source in the device system, and a stable ETC provides optimal device performance. The goal of this study was to reduce human labor in thermal characterization of packaging, which ultimately decreases risk of human error as well.

        \item \textbf{ML-based Classifiers for Predicting IC Yield in 3D Packaging:} ML algorithms can be used to predict results of final package device tests, with the goal of spotting bad dies coming from front-end fabs before they are packaged, using dataset obtained from early stage wafer fabrication testing for each production unit. This is especially critical for 3D IC packaging devices, where one bad die means the whole package fails. As a result, this leads to profit loss and puts extra burden on packaging engineers. Chen et al. \cite{Chen2017-xb} overcomes this issue by presenting a ML classifier model to maximize IC yields using a batch (to overcome imbalances in the dataset), online (the model is updated when a new dataset is received), and incremental learning (adapting the classifier to changes in data distribution as a result of non-linear manufacturing conditions) framework. This approach is tested on 3D flash memory chips using real data from an industry partner, where the ultimate goal is to separate bad dies, which undergo a lower cost testing and packaging process to be used in a low-end product. Predicting and isolating the good dies beforehand increases efficiency as the the packaging engineer is able to spend more time on packaging and memory testing the good dies, which also maximizes profits as the number of high-end products is increased. This developed algorithm resulted in a 3.4\% yield improvement in a 16 die stack case, which shows a promising ability to streamline the packaging process in a back-end manufacturing setting.

    \end{itemize}

\section{Conclusion}\label{sec8:discussion_and_conclusion}
Industry 4.0 has revolutionized the manufacturing industry with key capabilities in automation and AR/VR, which are currently not fully leveraged by the semiconductor industry. The developed future roadmaps paved the way for enhanced education, training, and employee experience through AR/VR as well as future streamlining of manufacturing processes through preventative/predictive maintenance, automated wafer/die quality control, and inspection of IC packaging for hardware assurance. Ultimately, this ensures a secure supply chain that upholds the U.S. in the global chip race amid the ongoing talent gap by employing the innovative potential of Industry 4.0 smart manufacturing.  All of these emerging paradigms possess promising outcomes in transforming the future of semiconductor manufacturing and production as a long-term solution by supporting CHIPS Act's ongoing workforce development efforts towards building a strong talent pipeline to expand domestic semiconductor manufacturing capacity. 
\section{Acknowledgment}\label{sec9:Acknowledgment}
The authors would like to thank ZEISS Research Microscopy Solutions (RMS) for providing their expertise in this research.

\bibliographystyle{IEEEtran}

\end{document}